\documentclass[11pt,twoside]{article}
\usepackage{CAGN2019}
\usepackage{graphicx}

\usepackage[T1]{fontenc} 

\usepackage{latexsym}
\usepackage{verbatim}

\usepackage{ifpdf}  
\ifpdf  
      \DeclareGraphicsExtensions{.pdf,.png,.jpg}  
\else  
      \DeclareGraphicsExtensions{.eps}  
\fi 

\begin{document}

\vskip 1.0cm
\markboth{J.~Garc\'{\i}a-Rojas et al.}{Latest advances in the abundance discrepancy problem in photoionized nebulae}
\pagestyle{myheadings}
%
%
\vspace*{0.5cm}
\parindent 0pt{Invited Review}


\vspace*{0.5cm}
\title{Latest advances in the abundance discrepancy problem in photoionized nebulae}

\author{J.~Garc\'{\i}a-Rojas$^{1,2}$, R.~Wesson$^3$, H.~M.~J. Boffin$^4$, D. Jones$^{1,2}$, R.~L.~M. Corradi$^{1,5}$, C. Esteban$^{1,2}$ and P. Rodr\'{\i}guez-Gil$^{1,2}$}
\affil{$^1$Inst. de Astrof\'{\i}sica de Canarias, E-38200, La Laguna, Tenerife, Spain\\
$^2$Dept. de Astrof\'{\i}sica, Univ. de La Laguna, E-38206, La Laguna, Tenerife, Spain\\
$^3$Department of Physics and Astronomy, University College London, Gower St, London WC1E 6BT, UK\\
$^4$European Southern Observatory, Karl-Schwarzschild-Str. 2, D-85738 Garching bei M\"unchen, Germany\\
$^5$6Gran Telescopio Canarias S.A., c/ Cuesta de San Jos\'e s/n, Bre\~na Baja, E-38712 Santa Cruz de Tenerife, Spain
}

\begin{abstract}
In this paper, we will focus on the advances made in the last few years regarding the abundance discrepancy problem in ionized nebulae. 
We will show the importance of collecting deep, high-quality data of H~{\sc ii} regions and planetary nebulae taken with the most advanced instruments 
attached to the largest ground-based telescopes. We will also present a sketch of some new scenarios proposed to explain the abundance discrepancy.


\end{abstract}

\section{Introduction}

In photoionized nebulae $-$both H~{\sc ii} regions and planetary nebulae (PNe)$-$ optical recombination lines (ORLs) provide abundance values that are systematically larger than those obtained
using collisionally excited lines (CELs). This is the well-known {\it abundance discrepancy} (AD) problem of nebular astrophysics, which has remained unresolved for more than seventy five years  \citep{wyse42}. This issue has obvious implications for the measurement of the chemical content of near and faraway galaxies, most often derived using CELs from their ionized interstellar medium. This abundance discrepancy is usually parametrized by the {\it Abundance Discprency Factor} (ADF) which is defined as the ratio of ORL and CEL abundances \citep{liuetal00}.  

Several scenarios have been proposed to explain this behavior. In the following we summarize some of the possible interpretations suggested to explain the abundance discrepancy: 
\begin{itemize}
\item \citet{peimbert67} was the first to propose the presence of temperature variations in the gas in order to explain the discrepancy between $T_e$([O~{\sc iii}]) and $T_e$(H~{\sc I}) derived from the Balmer jump. Subsequently, \citet{peimbertcostero69} proposed an scheme to correct the abundances computed from CELs for the presence of temperature inhomogeneities. \citet{torrespeimbertetal80} first suggested that the discrepancy between ORLs and CELs abundances could be explained if spatial temperature variations over the observed volume are considered. Possible causes to explain the presence of such temperature fluctuations in photoionized nebulae can be: mechanical energy injection by collisions \citep{peimbert95} or by conduction fronts \citep{maciejewskietal96}; high-density condensations \citep{viegasclegg94}; ionizing source variability \citep{binetteetal03, bautistaahmed18}; dust distribution inhomogeneities \citep{peimbertetal93, stasinskaszczerba01}; the presence of shadowed regions \citep{hugginsfrank06}; cosmic rays ionization of neutral clumps \citep{giammancobeckman05, zhangetal07} or the presence of dense X-ray irradiated regions \citep{ercolano09}.

\item The existence of chemical inhomogeneities in the gas was first proposed by \citet{torrespeimbertetal90} as a plausible mechanism to explain the abundance discrepancy. Later, \citet{liuetal00} proposed that metal-rich (i.~e. H-poor) inclusions could solve the abundance discrepancy problem; in this scenario, metal recombination lines are emitted in the metal-rich inclusions, where cooling has been enhanced, while CELs are emitted in the ``normal'' metallicity (H-rich) zones. Several authors have constructed two-phase photoionization models that sucessfully reproduce simultaneously ORLs and CELs emissions in H~{\sc ii} regions \citep{tsamispequignot05} and PNe \citep{yuanetal11}. However, the origin of such metal-rich inclusions has not been well established yet \citep[see][for a different scenarios proposed to explain the origin of such inclusions in PNe and H~{\sc ii} regions]{henneystasinska10, stasinskaetal07}.

\item \citet{nichollsetal12} proposed that a non Maxwell-Boltzmann $\kappa$ energy distribution of free electrons could explain the abundance discrepancies owing to the presence of a long tail of supra-thermal electrons that contribute to an increase in the intensity of the CELs at a given value of the kinetic temperature.  However, there is little theoretical \citep{mendozabautista14, ferlandetal16, drainekreisch18} or observational \citep{zhangetal16} support for this scenario in photoionized nebulae. In particular, \citet{ferlandetal16} have shown that the distance over which heating rates change are much longer than the distances over which supra-thermal electrons can travel, and that the timescale to thermalize these electrons is much shorter than the heating or cooling timescales. These estimates imply that supra-thermal electrons will have disappeared into the Maxwellian velocity distribution long before they affect the collisionally excited forbidden and RLs and  therefore, the electron velocity distribution will be closely thermal and the $\kappa$ formalism can be ruled out for these objects. Moreover, \citet{drainekreisch18} demonstrated analytically the departures from Maxwellian distribution have negligible effects on line ratios. These authors show that the electron energy distribution relaxes rapidly to a steady-state distribution that is very close to a Maxwellian.

\item Uncertainties in atomic data have also been traditionally claimed as a possible explanation for the abundance discrepancy problem. \citet{rodriguezgarciarojas10} found that the observed  temperature structure of Galactic H~{\sc ii} regions was easily reproduced with simple photoionization models with metallicities close to the ones implied by O CELs, and suggested that uncertainties in the ORLs recombination coeffcients could address this problem. \citet{storeyetal17} computed a new set of recombination coefficients for O$^{2+}$ accounting for departures of the local thermodynamic equilibrium, and extending the computations to low electron temperatures, $T_e$, where dielectronic recombination (DR) could be an important process. However, the new recombination coefficients applied to the available observations do not signifficantly change the obtained O$^{2+}$ abundances from ORLs. On the other hand, \citet{garnettdinerstein01} explored the possibility that high-temperature DR in a central hot bubble of the PN NGC\,6720 could enhance the ORLs strengths in the central parts of the nebula, finding that the increase of recombination rates due to DR at $T_e \approx 10^5$ K was not enough to overcome the increase in collisionally excited emission.  However, these authors conclude that  a better understanding of the effects of DR on recombination line strengths in general is urgently needed. Finally, we want to stress that  \citet{escalanteetal12} have shown that additional excitation processes of the ORLs, such as fluorescence excitation by starlight can solve the abundance discrepancy problem in low-ionized, low-ADF PNe such as IC\,418. However, this mechanism cannot explain the strong intensities of ORLs from $f$ or $g$ states, such as the widely observed C~{\sc ii} $\lambda$4267 line. 
\end{itemize}

\section{The abundance discrepancy in H~{\sc ii} regions}
\label{sec:hii}

Since the pioneering work by \citet{wyse42}, the abundance discrepancy problem has been studied in several Galactic \citep[see e. g.][]{garciarojasesteban07} and extragalactic \citep[see e.~g.][]{estebanetal14, toribiosanciprianoetal16} H~{\sc ii} regions. A compilation of ADF measurements made in both H~{\sc ii} regions and PNe can be found on Roger Wesson's webpage\footnote{https://www.nebulousresearch.org/adfs/}. In general, ADFs computed for H~{\sc ii} regions are in the range between 1.3 and 3, with an average value around 2 that is representative for most of the objects. Although the ADF in H~{\sc ii} regions can therefore be considered approximately constant,  \citet{toribiosanciprianoetal17} reported --for the first time-- an apparent dependence of both ADF(O$^{2+}$) and  ADF(C$^{2+}$) with metallicity (see their Figures~4 and 5), in the sense that the ADF seems to be higher as metallicity becomes lower. A similar result has also been reported by \citet{estebanetal18}, who found a correlation between the ADF with $T_e$ (see their Figure~4).  
Another test to explore the AD problem is to compare abundances derived from CELs and ORLs in H~{\sc ii} regions with those determined in young B and A-type supergiant stars of their vicinity. Results obtained by \citet{bresolinetal16} and \citet{toribiosanciprianoetal17} indicate that the ORL-based nebular metallicities agree with the stellar ones better than those from CELs in the high-metallicity regime, but that this situation reverses at low-metallicity values. Extending the baseline of the O/H ratios of H~{\sc ii} regions or emission-line galaxies with precisely derived ADF to lower metallicities would shed light on the dependence of the ADF with metallicity.

On the other hand, \citet{gusevaetal06, gusevaetal07} explored the behavior of $T_e$ derived from [O~{\sc iii}] CELs and from H~{\sc I} Balmer and Paschen jump, in a sample of low-metallicity H~{\sc ii} regions and emission-line galaxies finding no statistically significant differences. This result would rule out the presence of temperature fluctuations in such objects, and suggests that the ADF, if present, should be small. These results appear to go in the opposite direction to those obtained from the direct measurement of the ADF, and reinforce the need for better quality data to disentangle this puzzle.

Although the origin of the ADF in H~{\sc ii} regions is still unknown, it seems clear that several actors are contributing to it. \citet{mesadelgadoetal08} used long-slit spectroscopy to cover several morphological structures in the Orion nebula, such as Herbig-Haro (HH) objects and protoplanetary disks (proplyds), finding prominent spikes of $T_e$([N~{\sc ii}]) and $n_e$. They also found a significant enhancement of the ADF  at the locations of the most conspicuous HH objects. \citet{nunezdiazetal12}, using integral field spectroscopy, discovered a narrow arc of high $T_e$([{N~\sc ii}]) at the head of the bow-shock of HH\,204 in the Orion nebula. These arcs have been predicted by models of photoionized HH jets of \citet{ragareipurth04}, which are produced by shock heating at the leading working surface of the flow, just preceding the high-density compressed zone. \citet{mesadelgadoetal09}, using high-spectral resolution spectra at the head of HH\,202-S, and were able to separate the spectrum of the high-velocity flow from the ambient ionized gas. These authors found large density differences between both components, as well as evidences of dust destruction and abundance anomalies in the gas flow component (i.~e.  CEL abundances were 0.2 dex lower than in the ambient gas, while ORL abundances remained unchanged), indicating that these gas flows play a role in the AD problem.

Additionally,\citet{tsamisetal11} and  \citet{mesadelgadoetal12}, using integral field spectroscopy of fields in the Orion nebula containing proplyds, found that collisional de-excitation has a crucial influence on the line fluxes in the proplyds, owing to the high electron densities ($n_e$$\sim$$10^5-10^6$ cm$^{-3}$) in these objects. They also found that the use of suitable density diagnostics and a proper background substraction returned an ADF(O$^{2+}$)$\approx 1$ for the intrinsic spectra of the proplyds, and therefore concluded that the presence of high-density ionized gas can severely affect the abundances determined from CELs.  

\section{The abundance discrepancy in planetary nebulae}
\label{sec:pne}

In the last years there has been multiple studies on the AD problem in PNe. A compilation of the results obtained for PNe has been made by Roger Wesson on his webpage (see Section~\ref{sec:hii}). ADFs in PNe show a behavior that is remarkably different to that of H~{\sc ii} regions in the sense that, although the bulk of PNe shows an ADF distribution with values that are usually between 1.5 and 3, this distribution has a significant tail extending to much larger values. \citet{garciarojasesteban07} proposed that, at least for the most extreme ADF PNe, the origin of the abundance discrepancy should be different to that of the ``normal'' ADF PNe and H~{\sc ii} regions.

Focusing on the physical conditions derived using different diagnostics, \citet{zhangetal04} derived $T_e$ and $n_e$ using the nebular hydrogen recombination spectrum, as well as $T_e$ obtained from the ratio of the fine-structure far-IR [O~{\sc iii}] 52, 88 $\mu$m to the nebular [O~{\sc iii}] 4959 \AA\  line, and compared the results with those obtained with the traditional nebular-to-auroral [O~{\sc iii}] 4959/4363 ratio, finding that temperature and density fluctuations are generally present  within PNe. They also found that $T_e$ derived from the far-IR to optical diagnostic was generally higher than that derived from the optical [O~{\sc iii}] lines, which was attributed to the existence of dense clumps in the PNe. This discrepancy in $T_e$ was found to be anti-correlated with $n_e$, suggesting that it is related to nebular evolution. 

Several authors have found evidence of the existence of two components with different physical conditions coexisting in PNe. \citet{storeysochi14} studied the continuum emission spectrum of PN Hf\,2-2 and found that the model which best fit the observations was one comprising two components with very different temperatures. \citet{zhangetal05} found that $T_e$ derived from He~{\sc i} lines in PNe were lower than those derived from the H~{\sc i} Balmer decrement, which is at odds with what is expected from the temperature fluctuations scenario, but consistent with a two-abundance model. \citet{storeyetal17} derived new recombination coefficients for O~{\sc ii} ORLs, using an intermediate coupling treatment that fully accounts for the dependence of the distribution of the population among the ground levels of O$^{2+}$ on $n_e$ and $T_e$, and extending the computations down to low $T_e$ to account for important effects owing to DR. These authors found that using a pure O~{\sc ii} ORLs $T_e$ diagnostic it is clear that, at least for the extreme ADF PNe, the emission seems to come from a very low $T_e$ region. 
On the other hand, other authors have a different point of view. \citet{peimbertetal14} found for a sample of $\sim$20 PNe, that the average temperatures derived from H, He and O lines were consistent with a chemically homogeneous gas with temperature fluctuations. However, they could not rule out chemical inhomogeneities for 4 objects of their sample.

Another approach to search for evidence of different gas components is PNe is to study the kinematics of the gas. \citet{richeretal13}, from a spatially-resolved high-resolution UVES spectra of the PN NGC\,7009 found that the kinematics of C~{\sc ii}, N~{\sc ii}, O~{\sc ii}, and Ne~{\sc ii} ORLs did not coincide with the kinematics of the [O~{\sc iii}] and [Ne~{\sc iii}] CELs, indicating that there is an additional plasma component where the ORLs are emitted which arises from a different volume to that giving rise of the CELs. Moreover, \citet{richeretal19} studied the temperature structure of NGC\,7009 using the same data set and found large differences between O~{\sc ii}-based $T_e$ and CELs-based $T_e$ (by more than 6000 K) in some parts of the nebula, strengthening the hypothesis of the existence of two plasma components in NGC\,7009. Similar results have been found by \citet{penaetal17}. These authors studied the expansion velocities ($V_{exp}$) in 14 PNe from O$^{2+}$ ORLs and CELs and found that, in general, if the gas is in ionization equilibrium, ORLs have lower $V_{exp}$ than CELs for a given ion, which means that ORLs are emitted by a gas component located closer to the central star. Finally, from a statistical study of the kinematics of the C~{\sc ii} 6578 permitted line in 83 lines of sight in 76 PNe, \citet{richeretal17} found that the kinematics of this line was not that expected if the line arose from the recombination of C$^{2+}$ ions or the fluorescence of C$^+$ ions in ionization equilibrium in a chemically homogeneous nebula, but instead its kinematics were appropriate for a more internal volume than expected.

\subsection{The link between close binary central stars and large ADFs }
\label{subsec:bin}

\citet{wessonetal03} proposed for the first time that PNe exhibiting particularly large ADFs, including those `born-again' PNe, could be related to the phenomenon of novae and, therefore, host a central binary star. \citet{liuetal06} support this scenario as they found an extreme ADF ($\sim$70) in the PN Hf\,2-2 which is known to host a close binary system that has been suggested to result from a common envelope (CE) event.  Inspired by these findings, \citet{corradietal15} analysed the nebular spectra of 3 PNe (Abell\,46, Abell\,63, and Ou\,5) known to host post-CE close central binary stars, and they found extreme ADFs (ADF$>$10) in two of them and a very high value (ADF$\sim$10) in the third one. Furthermore, spatially-resolved analysis revealed that the ADFs were centrally peaked  (reaching up to 300 in the central part of Abell\,46). These results along with those obtained by \citet{jonesetal16} for NGC\,6778, another PN with a post-CE binary central star and a high ADF$\sim$20, confirm that the ADF is strongly centrally peaked in these objects. However, this result is not restricted only to PNe with high-ADFs and known binary central stars, and can be found in low-ADF PNe \citep[see][and Section~\ref{sec:pne}]{garnettdinerstein01}. In an attempt to shed light on this issue, \citet{garciarojasetal16} obtained unprecedented tunable filter imaging of the emission coming from O~{\sc ii} ORLs and compared it with narrow-band imaging in the [O~{\sc iii}] $\lambda$5007 CEL in NGC\,6778; these authors discovered that the spatial distribution of the O$^{2+}$ ions producing the O~{\sc ii} ORLs does not match that of the O$^{2+}$ ions emitting in the [O~{\sc iii}] CELs. This is a clear indication of the presence of two separate plasmas that are not well mixed, perhaps because they were produced in distinct ejection events. These results confirm the hypothesis that large ADFs in PNe and low-to-moderate ADFs in H~{\sc ii} regions and PNe probably have different physical origins \citep{garciarojasesteban07}. 

In a recent paper, \citet{wessonetal18} give extra support to the link between large ADFs and binarity of the central star from the analysis of a larger sample of PNe with known post-CE central stars, which present, in all the cases ADFs ranging from 10 to 80. These authors have also confirmed an anti-correlation between abundance discrepancies and nebular $n_e$. On the other hand, central star binarity seems to be a necessary but not sufficient condition for having a large ADF. \citet{manicketal15} discovered significant periodic variability of the central star of PN NGC\,5189 that revealed a close binary with an orbital period of $\sim$4.04 days. However, \citet{garciarojasetal13} found an ADF of 1.6 for this object. \citet{sowickaetal17} computed and ADF of 1.75 for the PN IC\,4776 and reported that the central star was a binary system with a period of $\sim$9 days, although in this case, the period is poorly constrained. Additionally, \citet{corradietal15} reported that the lack of detection of O~{\sc ii} ORLs in deep spectra of PNe sets a low upper limit to the ADF; this is the case of  the Necklace nebula, a well known post-CE object with an orbital period of its central system of 1.16 days, where no ORLs were detected in very deep spectra obtained by \citet{corradietal11}. Taking all these results and the available observations into account, \citet{wessonetal18} concluded that all nebulae with a binary central star with a period of less than 1.15 days have large ADFs (ADF$\geq$10 and low electron densities ($n_e < 1000$ cm$^{-3}$). Moreover, these authors also reported for the first time that [O~{\sc ii}] density diagnostic lines can be strongly enhanced by recombination excitation, while  [S~{\sc ii}] lines are not, suggesting that in the PNe where $n_e$([O~{\sc ii}]) signifficantly exceeds $n_e$([S~{\sc ii}]) this can be used as an additional diagnostic to infer an extreme-ADF PN and hence, the presence of a close binary star, even if ORLs lines are not detected. Finally, the evidence found in their study led \citet{wessonetal18} to support the idea that extreme ADFs are caused by nova-like eruptions from the central system, occuring soon after the CE phase, which ejects material depleted in H and enhanced in C, N, O and Ne, but not in third-row elements.

\begin{figure}[ht]  
\begin{center}
\hspace{0.25cm}
\includegraphics[width=9cm]{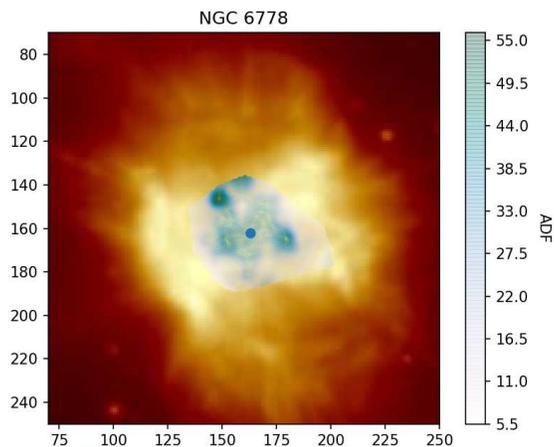}
\caption{Preliminary results of the ADF spatial distribution in the PN NGC\,6778 from our MUSE data. Background image is the H$\alpha$ emission map. ADF map covers only the zones were O~{\sc ii} ORLs were detected. It can be seen that the ADF peaks in regions close to the center of the nebula, reaching values higher than 50.}
\label{fig1}
\end{center}
\end{figure}

Our group has taken advantage of the high spatial-resolution of the 2D spectrograph MUSE at VLT to confirm that in several high-ADF PNe, the ORL emission is more centrally concentrated than the emission of the strongest [{O~{\sc iii}}] $\lambda$4959 CEL \citep[see Fig.~1 of][]{garciarojasetal17}. As pointed out above, this is very frequently related with the presence of a confirmed post-CE binary central star, but sometimes it can be a signal of binarity of a still unconfirmed binary central star.  This result, together with the presence of a high-ADF, can be a signal of binarity of a still unconfirmed binary central star.
In Fig.~\ref{fig1} we also show a preliminary computation of the spatial distribution of the ADF in NGC\,6778, showing that the value clearly peaks in the central regions of the PN. A detailed analysis of the spatial distribution of the physical conditions and ionic abundances, from both CELs and ORLs line ratios is underway. 

\begin{figure}[ht]  
\begin{center}
\hspace{0.25cm}
\includegraphics[width=9cm]{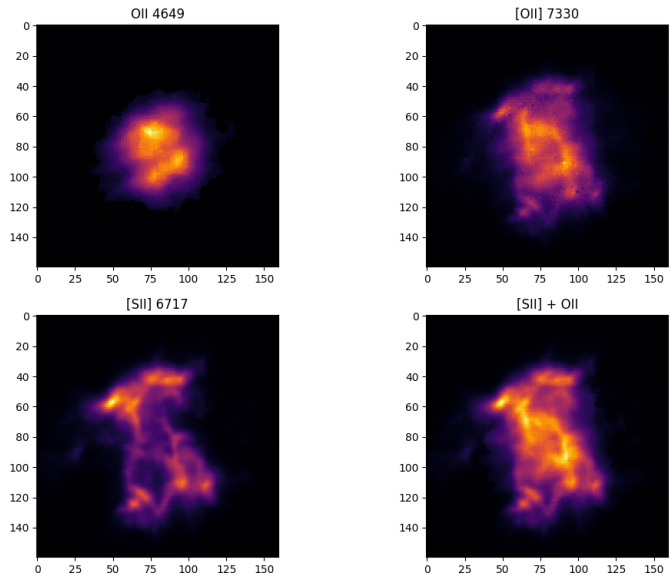}
\caption{MUSE emission line maps of PN NGC\,6778. From left to right and upper to down: O~{\sc ii} $\lambda$4649+50 ORL, [O~{\sc ii}] $\lambda$7330 CEL, [S~{\sc ii}] CEL and a composite image of [S~{\sc ii}]+O~{\sc ii} emission maps (see text).}
\label{fig2}
\end{center}
\end{figure}

Another possibility to explain the behavior of the high-ADF PNe has been recently proposed by \citet{alidopita19}. Using IFU data of the PN PB\,4, these authors found strong O~{\sc ii}  ORL emission lines and a similar spatial distribution of O~{\sc ii} $\lambda$4649 ORL and [O~{\sc ii}] $\lambda\lambda$7320 CEL and proposed that the O~{\sc ii} ORLs do not arise from recombination at all, but from fluorescent pumping of excited states in the O$^+$ ion caused by the extreme UV continuum of an interacting binary central star. We investigated the behavior of  O~{\sc ii} ORLs and [O~{\sc ii}] CELs in our MUSE data of NGC\,6778 and found that [O~{\sc ii}] emission is much more extended than  O~{\sc ii} emission (see Fig~\ref{fig2}) and that the possible similarities found in the emission distribution comes from the fact that [O~{\sc ii}] $\lambda$7320+30 lines are strongly affected by recombination contribution. 
Strong recombination line enhancements are seen in second-row elements but not third-row ones \citep[][and references therein]{wessonetal18}
If O$^{2+}$ is strongly affected by recombination but S$^{2+}$ is not, then, as the two ions have similar ionisation potentials, the spatial distribution of  of [O~{\sc ii}] should consist of a collisional component resembling [S~{\sc ii}], and a recombination component resembling O~{\sc ii} ORLs.
In Fig.~\ref{fig2}, we compare a map of NGC\,6778 in the [O~{\sc ii}] $\lambda$7330 CEL with with a composite image created by adding a map of [S~{\sc ii}] emission (scaled to the same peak flux as the  [O~{\sc ii}] image) and a map of the O~{\sc ii} $\lambda$4649+50 ORL. This composite image indeed strongly resembles the [O~{\sc ii}] image. Moreover, if fluorescent pumping is affecting the emission of O~{\sc ii} ORLs, it should also affect in a similar way the emission of C/N/Ne ORLs that are also strongly enhanced in extreme ADF PNe. Given the different atomic configurations for these elements, it is hard to reconcile the scenario proposed by \cite{alidopita19} with observations. 

\section{Conclusions}
\label{sec:conclu}

In the last 15 years significant advances have been made in the field of the AD problem. Deep high-quality spectra have revealed to be crucial to advance understanding of this problem. Close binary evolution has been shown to be linked with the high ADFs found in PNe, opening a research line encouraging observational and theoretical astronomers to look for additional probes that can help to understand the CE process in close binary stars and the AD phenomenon. In particular, high-spatial and -spectral resolution 2D spectroscopy with MEGARA at GTC and FLAMES at VLT will be crucial to obtain detailed information about the spatial distribution and kinematics of the two gas components. Finally, the combination of these data with more refined models of the CE process and 3D photoionization models have the potential to revolutionize the way we can tackle this fundamental problem.

On the other hand, the solution to AD problem in H~{\sc ii} regions and in the bulk of PNe is still far from being understood. Additional spatially-resolved observations with high-spectral resolution covering morphological structures in H~{\sc ii} regions and PNe like high-velocity ionized flows of gas or proplyds would, for sure, give precious information to study the impact of these structures in the integrated spectrum of the nebulae. 

\acknowledgments 
JGR acknowledges support from an Advanced Fellowship from the Severo Ochoa excellence program (SEV-2015-0548). This research has been supported by the Spanish Ministry of Education, Science and Sports under the grants AYA2017-83383-P.  RW was supported by European Research Grant SNDUST 694520. CE acknowledges support from the project AYA2015-65205-P 

\bibliographystyle{aaabib}
\bibliography{examplebib}

\end{document}